\documentclass[conference]{IEEEtran}
\ifCLASSINFOpdf
\else
\fi
\usepackage{graphicx}

\begin{document}
%
\title{NFT MARKETPLACE}

\author{
\IEEEauthorblockN{Piyush Batra}
\IEEEauthorblockA{\textit{Department of Computing Science} \\
\textit{University of Alberta}\\
pbatra2@ualberta.ca}
\and
\IEEEauthorblockN{Gagan Raj Singh}
\IEEEauthorblockA{\textit{Department of Computing Science} \\
\textit{University of Alberta}\\
grsingh@ualberta.ca }
\and
\IEEEauthorblockN{Ritik Gandhi}
\IEEEauthorblockA{\textit{Department of Computing Science} \\
\textit{University of Alberta}\\
rgandhi1@ualberta.ca }}


%


\maketitle

\begin{abstract}
In an increasingly digitized world, the secure management and trade of digital assets has become a pressing issue. This project aims to address this challenge by developing a decentralized application (dApp) that leverages blockchain technology and deep learning models to provide secure and efficient digital asset management, with a focus on NFTs. The dApp includes features such as secure wallet connections, NFT image generation, minting, marketplace, and profile management. The back-end of the dApp is implemented using the Goerli testnet with Solidity-based smart contracts, while IPFS and ReactJS/EtherJS are used for decentralized storage and front-end development, respectively. Additionally, the OpenAI API is integrated to generate unique NFT images based on user input. The project demonstrates the practical application of blockchain technology and deep learning models in developing dApps for secure and decentralized digital asset management. Overall, the project contributes to the ongoing research on blockchain-based solutions for secure digital asset management, while highlighting the potential of blockchain and deep learning technologies to transform the way we manage and trade digital assets.
\end{abstract}


%
\IEEEpeerreviewmaketitle

\section{Introduction}
Blockchain technology has created new possibilities for managing and trading digital assets, with non-fungible tokens (NFTs) being one of the most intriguing applications of this technology. However, ensuring the security of NFTs is a major challenge, particularly concerning safeguarding users' private keys. The purpose of this project is to address these challenges by developing a decentralized application (dApp) \cite{dapp} named NFT Marketplace, which offers a safe and user-friendly platform for NFT management.

This project aims to tackle several fundamental questions regarding the secure management and trading of NFTs. How can users connect their cryptocurrency wallets securely to a dApp, allowing them to interact with NFTs without exposing their private keys? How can unique NFTs be generated utilizing deep learning models and then uploaded to decentralized storage? How can NFTs be traded securely and transparently on a blockchain-based marketplace? And how can users manage and view their NFT collections simply and intuitively?

To address these questions, we have designed and implemented the NFT Marketplace. This platform enables users to securely connect their wallets to the dApp, create unique NFTs using deep learning models, and manage their NFT collections. Smart contracts written in Solidity \cite{solidity} for secure and transparent asset transfer on the blockchain were used with the Goerli testnet \cite{goerli}. Additionally, we used the OpenAI API \cite{openai} to generate unique NFT images based on user input, which were then uploaded to decentralized storage on IPFS \cite{IPFS}.

In the subsequent sections of this report, we will describe the design and implementation of our dApp and present results from our testing and evaluation of the dApp's performance, usability, and discuss the limitations of our approach and future directions for research and development in this area. Furthermore, we will present the results of the performance evaluation of our dApp in terms of the time taken for API calls during the NFT generation and minting process. We will also provide details of the usability study conducted to assess the user-friendliness of our dApp.

\section{Design and Implementation}
The development process of our NFT Marketplace dApp involved several key stages, including designing the user interface, implementing smart contracts on the blockchain, integrating with third-party APIs, and conducting testing and evaluation. In this section, we provide a detailed overview of each stage, including the tools and technologies used, the challenges encountered, and the solutions developed. Additionally, we describe the methodologies used for connecting users' wallets securely, generating unique NFTs using deep learning models, minting NFTs on the blockchain, developing the marketplace for trading NFTs, and creating user profiles to manage NFT collections. By providing a comprehensive account of our development process, we aim to provide insights and best practices for building user-friendly and secure dApps for managing and trading digital assets on the blockchain.

\subsection{Wallet Connect}
The Wallet Connect feature is a critical component of the NFT Marketplace dApp as it enables users to securely connect their cryptocurrency wallets to the dApp without exposing their private keys. This is achieved through the use of Metamask \cite{metamask}, a popular browser extension that acts as a bridge between the user's browser and their cryptocurrency wallet.

When a user accesses the NFT Marketplace dApp, they are prompted to connect their wallet through Metamask. Once connected, the user can interact with the dApp's features, such as generating and minting NFTs, without having to manually input their private keys. This eliminates the risk of key theft or unauthorized access to the user's digital assets.

To enable secure and user-friendly interaction with the NFT Marketplace dApp, we utilized React.js \cite{react} with Ether.js\cite{etherjs}, a popular library for interacting with Ethereum-based networks. This library allowed us to connect the dApp with users' cryptocurrency wallets through the Metamask browser extension. The Ether.js library provided a simple and intuitive API for sending and receiving data from the blockchain, making it easy for us to integrate the dApp with the Ethereum \cite{eth} network.

\subsection{NFT Generation}
NFT generation is a crucial aspect of our application that involves using a deep learning model to create unique NFT images based on user input. To implement this feature, we utilized the OpenAI API \cite{openai}, which provides access to a pre-trained DALL·E model capable of generating images from text prompts.

\begin{figure}[!h]
    \centering
    \includegraphics[width=0.5\textwidth]{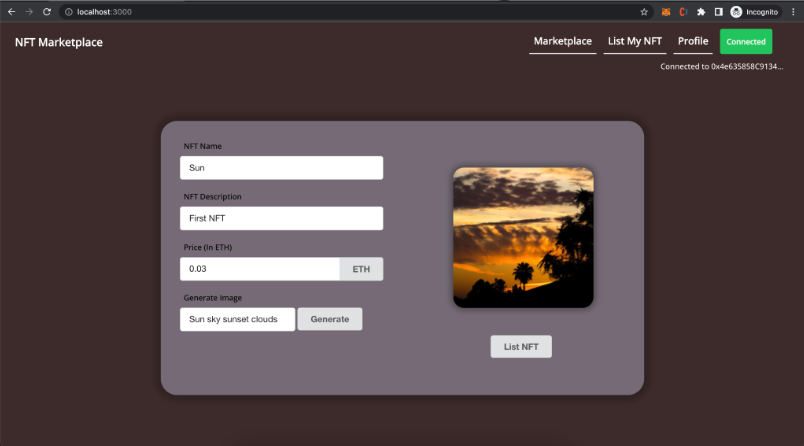}
    \caption{NFT Generation Page}
    \label{fig: UIpage1}
\end{figure} 

To generate an NFT, a user inputs a set of seed words or phrases that describe the desired characteristics of the NFT. These inputs are passed through our backend Express\cite{express} server that communicates with the OpenAI API, which generates a unique image using the DALL·E model. This model is trained on a massive dataset of images and text prompts, which enables it to generate high-quality and diverse NFT images based on user input. The use of a deep learning model also allows for the creation of personalized and unique NFTs that cannot be easily replicated.

The integration of blockchain technology and deep learning in the NFT generation process allows for the creation of valuable and distinctive digital assets. By leveraging the power of these technologies, the NFT Marketplace dApp provides a unique and exciting way for users to create, own, and trade NFTs.

\subsection{NFT Minting}
NFT minting feature enables users to create and sell unique digital assets on the blockchain. To implement this feature, we utilized a combination of the OpenAI API, IPFS, and the Ethereum \cite{eth} blockchain.

To mint an NFT, the user first generates an image using the OpenAI API. Once the image is generated, the user clicks the "List NFT" button, which initiates the minting process. At this point, a JSON object is created containing the NFT's name, description, price, and the image URL provided by the OpenAI API. This JSON file is then pinned to IPFS using the Pinata API's pinJSONToIPFS method \cite{pinata}, which returns an IPFS hash that serves as the metadata URI for the NFT.

\begin{figure}[!h]
    \centering
    \includegraphics[width=0.5\textwidth]{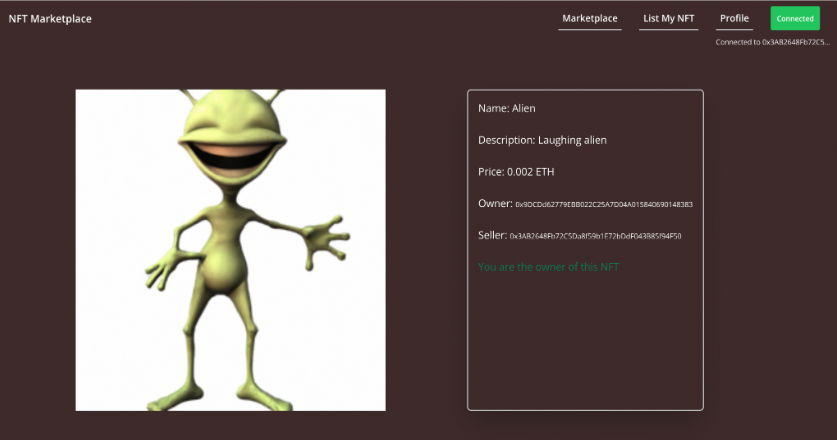}
    \caption{NFT Description Page}
    \label{fig: NFTPage}
\end{figure} 

The mint function in the smart contract is then called, which creates a new token on the Ethereum blockchain and sets its URI to the IPFS hash returned by the Pinata API. The smart contract also specifies the characteristics of the token, such as its name, symbol, and the total number of tokens in existence. Once the function is executed, the NFT is officially minted and can be viewed in the Marketplace. The metadata URI stored on IPFS allows anyone to view the NFT's name, description, price, and image. The token ID on the blockchain ensures that the NFT is unique and can be transferred between users.

\begin{figure}[!h]
    \centering
    \includegraphics[width=0.5\textwidth]{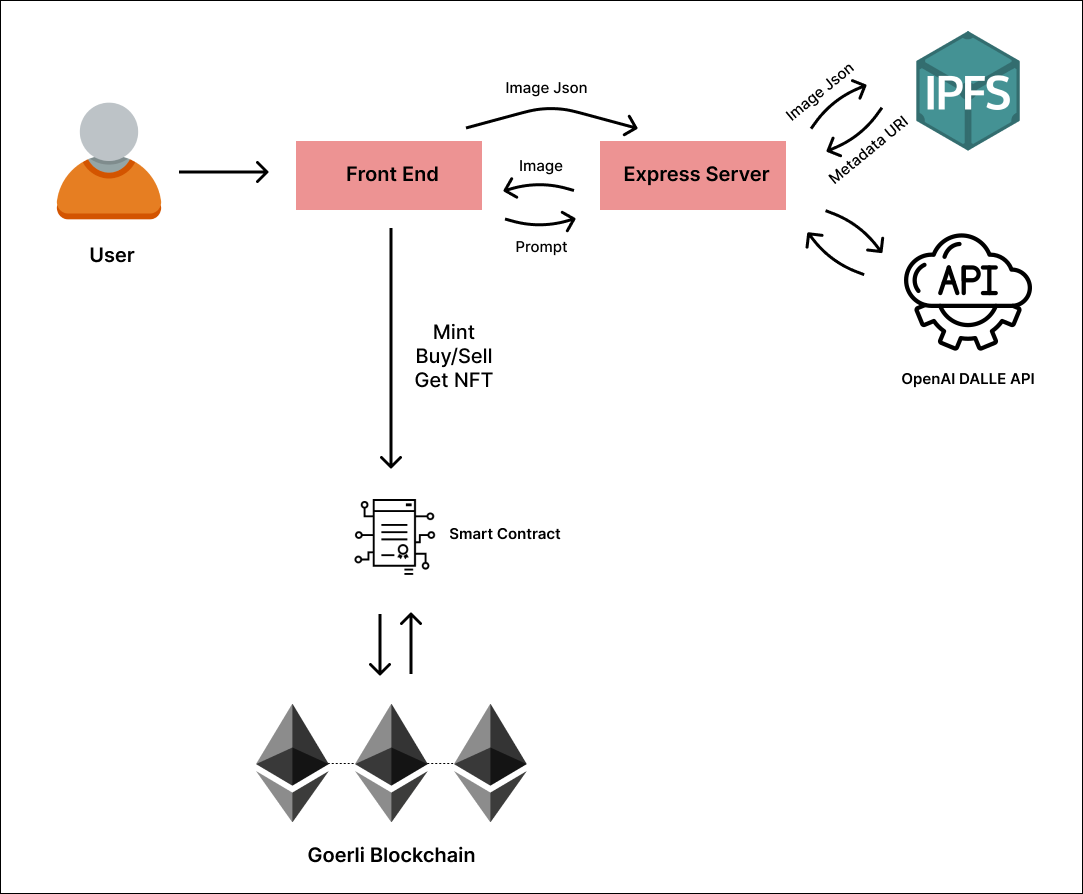}
    \caption{Workflow of our application}
    \label{fig: Workflow}
\end{figure} 

By linking the NFT token to the IPFS hash, we ensure that the NFT's metadata is decentralized and tamper-proof. Additionally, the use of a smart contract on the Ethereum blockchain enables users to securely and transparently transfer ownership of their NFTs, as ownership information is stored immutably on the blockchain.

\subsection{Marketplace}
The Marketplace functionality of this application enables users to browse and purchase NFTs created by other users. The feature is built on the Ethereum blockchain, utilizing the Goerli testnet\cite{goerli}, and allows users to interact with the smart contract that handles the creation, storage, and transfer of NFTs.

The marketplace displays all available NFTs stored on the contract, allowing users to browse through them and view their associated metadata, such as name, description, and price. The metadata is stored on IPFS, ensuring decentralization and tamper-proofing.

To purchase an NFT, a user clicks on the "Buy" button associated with the desired NFT, triggering a call to the smart contract's buyToken function. The function transfers ownership of the NFT to the user and updates the smart contract's storage to reflect the new ownership. The transaction is then recorded on the blockchain, providing an immutable record of the transfer.

\begin{figure}[!h]
    \centering
    \includegraphics[width=0.5\textwidth]{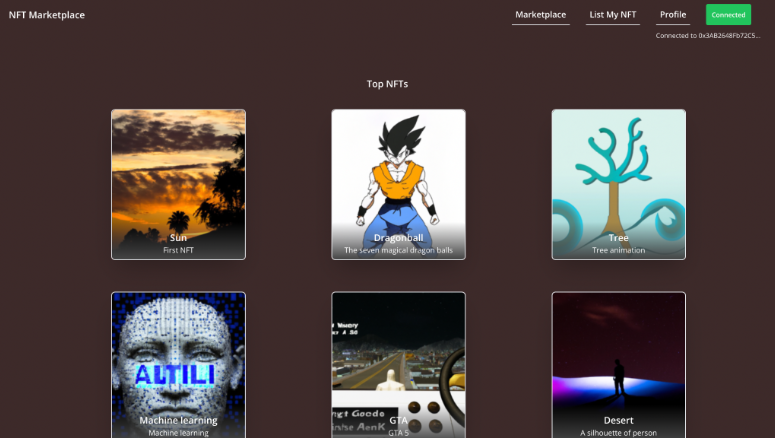}
    \caption{NFT Marketplace Page}
    \label{fig: UIpage2}
\end{figure} 

The dApp is designed with a user-friendly interface that includes clear instructions and intuitive buttons for browsing, purchasing, and selling NFTs. The use of blockchain technology and IPFS ensures the security and integrity of the NFT transactions and metadata, providing a seamless user experience.

\subsection{User Profile}
The UserProfile page is a key component of the NFT Marketplace dApp, providing users with an overview of their account's NFT ownership and value. The Ethereum blockchain's smart contract capabilities are integrated into the dApp's interface to implement this feature.

\begin{figure}[!h]
    \centering
    \includegraphics[width=0.5\textwidth]{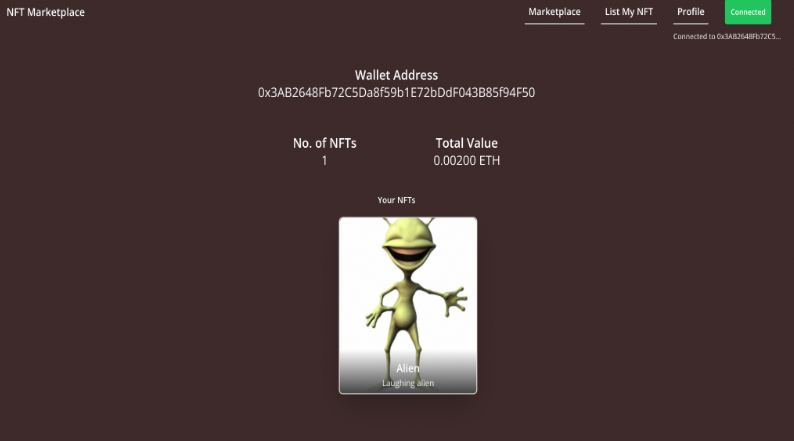}
    \caption{UserProfile Page}
    \label{fig: UIpage3}
\end{figure} 

When a user accesses their UserProfile page, the dApp retrieves data that relates to the user's account. This data includes the user's wallet address, the total number of NFTs owned, and the total value of the NFTs in the user's possession. The dApp displays this information in a user-friendly format on the UserProfile page. This page page also displays all NFTs owned by the user. The dApp queries the Ethereum blockchain's smart contract to retrieve the list of NFTs owned and fetches the NFT metadata from IPFS using the metadata URI associated with each NFT token ID. The information is displayed in a visually appealing and easy-to-navigate format.

The integration of smart contracts into the UserProfile page enables users to securely and transparently view their NFT ownership and value on the blockchain.

\subsection{Team Work and Contributions}
Our project was a collaborative effort all three among team members. Each team member was assigned different tasks based on their strengths and areas of expertise. We followed an agile methodology to ensure that each task was completed within the given timeline.

The table below shows the breakdown of tasks assigned to each team member:

\begin{table}[!h]
\centering
\begin{center}
\resizebox{\columnwidth}{!}{%
\begin{tabular}{|l|l|llllll}
\cline{1-2}
\textbf{Task}                                & \textbf{Assigned To}                      &  &  &  &  &  &  \\ \cline{1-2}
Define Requirements                          & \multicolumn{1}{r|}{All} &  &  &  &  &  &  \\ \cline{1-2}
NFT generation architecture              & \multicolumn{1}{r|}{Gagan, Piyush}      &  &  &  &  &  &  \\ \cline{1-2}
Client side of the web app                   & \multicolumn{1}{r|}{Gagan, Ritik}         &  &  &  &  &  &  \\ \cline{1-2}
Integrating the web app with web3.js libraries & \multicolumn{1}{r|}{Piyush, Ritik}      &  &  &  &  &  &\\ \cline{1-2}
Implementing the minting process               & \multicolumn{1}{r|}{Gagan, Piyush}      &  &  &  &  &  &  \\ \cline{1-2}
Connect the generated objects to the blockchain & \multicolumn{1}{r|}{Piyush, Ritik}        &  &  &  &  &  &  \\ \cline{1-2}
Testing and deploying the contracts          & \multicolumn{1}{r|}{Gagan, Ritik}       &  &  &  &  &  &  \\ \cline{1-2}
Testing the web app and conducting usability study          & \multicolumn{1}{r|}{Gagan, Piyush}        &  &  &  &  &  &  \\ \cline{1-2}
\end{tabular}%
}
\end{center}
\caption{Team Work}
\label{tab:my-table}
\end{table}

\section{Project status}

The project has been implemented with the intended functionality of generating and minting NFTs on the Ethereum blockchain. All requirements specified in the project scope were successfully met, and the team has been able to deliver a fully functioning web application that allows users to generate and mint their own NFTs.

Throughout the project, the team has encountered several technical difficulties, primarily in the integration of the web app with the web3.js libraries, which led to a delay in the project timeline. However, the team was able to overcome these issues by seeking help from online communities and resources.

In conclusion, the project has achieved its intended goals and has been completed successfully, albeit with some minor setbacks. The team has gained valuable experience in developing a blockchain-based web application, and the knowledge gained can be applied in future projects related to blockchain technology.

\section{Results}
In this section, we present the results of the performance and usability study of our project. The performance test involved measuring the time it took to generate NFT images using OpenAI's API and mint NFTs using smart contract transactions on the Goerli testnet \cite{goerli}. For the usability study, we surveyed ten users around the campus to evaluate the ease of use, clarity, and visual appeal of the marketplace, as well as the buying process and overall user experience. We present the results of the usability study in the form of bar graphs to illustrate the responses of the participants.

\subsection{Performance Evaluation}
The performance test involved generating NFT images using OpenAI's DALL·E  models and minting NFTs using smart contract transactions on the Goerli testnet \cite{goerli}. We plotted two graphs to visualize the results of the performance test. The x-axis of the graphs shows the unique requests made, and the y-axis shows the time taken for each request. 

\begin{figure}[!h]
    \centering
    \includegraphics[width=0.5\textwidth]{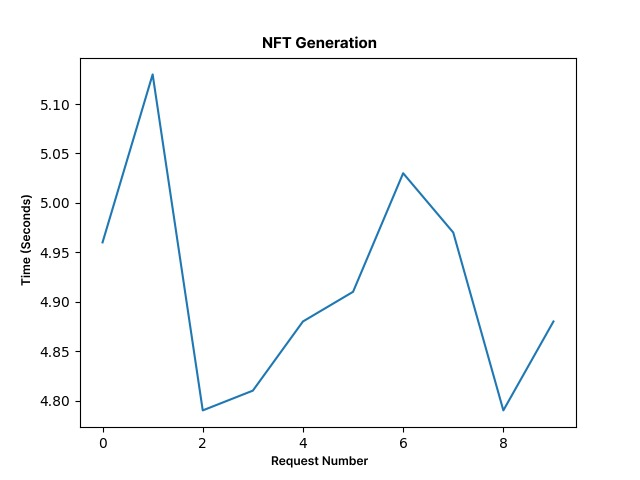}
    \caption{NFT generation performance}
    \label{fig: Generation}
\end{figure} 

For image generation, the results showed that the average time taken to generate an image was around 4.9 seconds, with the maximum time taken being 5.15 seconds. 

\begin{figure}[!h]
    \centering
    \includegraphics[width=0.5\textwidth]{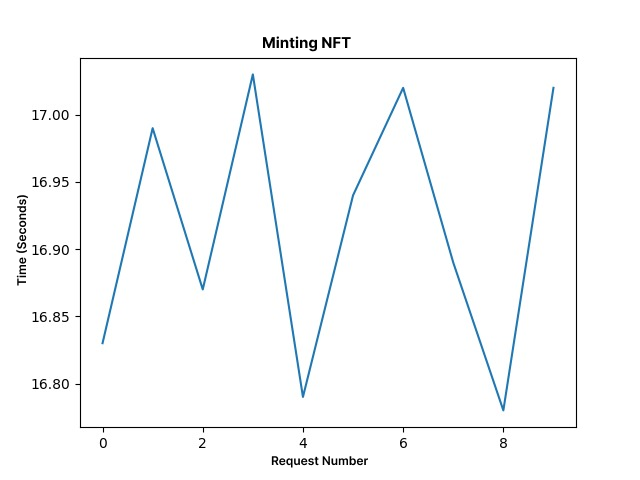}
    \caption{NFT minting performace}
    \label{fig: Minting}
\end{figure} 

For minting, the average time taken for minting an NFT was around 16.9 seconds, with the maximum time taken being 17.05 seconds. The results indicate that the performance of the NFT marketplace prototype is acceptable, but there is room for improvement, especially for the minting process, which takes longer than the image generation process. One of the factors contributing to this delay is the use of the Goerli testnet \cite{goerli}, which is not as fast as the Ethereum mainnet.

\subsection{Usability Study}

We conducted a usability study with 10 users from around the campus, asking them to rate their experience using the marketplace. On a scale of 1 to 5, 5 users found the marketplace easy to navigate and use, while the remaining 5 users rated it as very easy to use.
\begin{figure}[!h]
    \centering
    \includegraphics[width=0.5\textwidth]{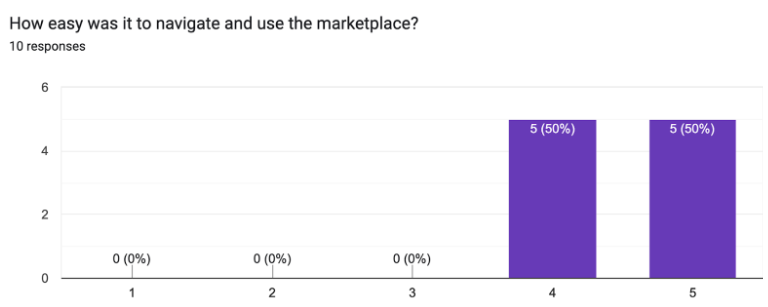}
    \caption{Usability study Q1 results}
    \label{fig: Q1}
\end{figure} 

 Similarly, 6 users found the image generation process clear and easy to use, while 4 users rated it as very easy to use.

\begin{figure}[!h]
    \centering
    \includegraphics[width=0.5\textwidth]{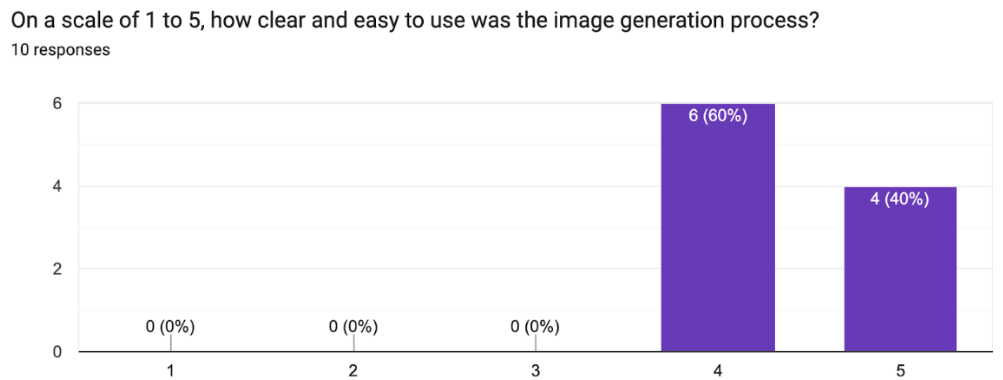}
    \caption{Usability study Q2 results}
    \label{fig: Q2}
\end{figure} 

 When it comes to minting an NFT, 5 users rated it as very easy, 4 users found it easy, and 1 user found it slightly difficult. 

\begin{figure}[!h]
    \centering
    \includegraphics[width=0.5\textwidth]{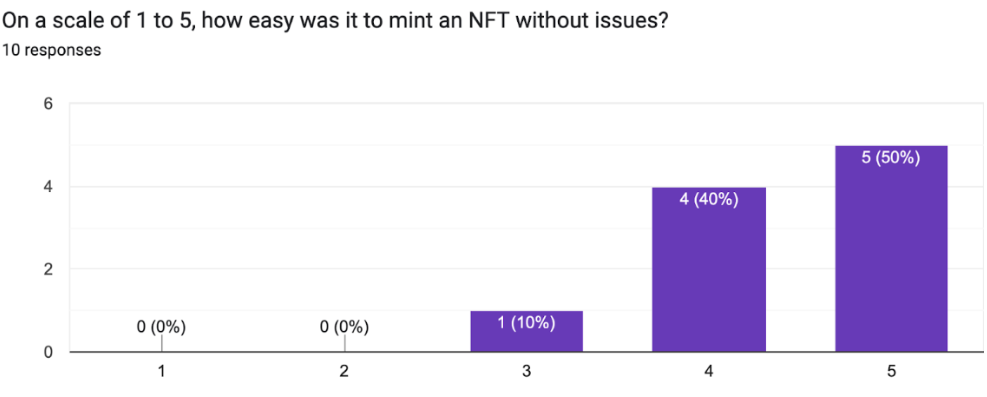}
    \caption{Usability study Q3 results}
    \label{fig: Q3}
\end{figure} 

In terms of the visual appeal of the NFT marketplace, 4 users rated it as very appealing, 3 users found it somewhat appealing, and 3 users found it to be somewhat unappealing.

\begin{figure}[!h]
    \centering
    \includegraphics[width=0.5\textwidth]{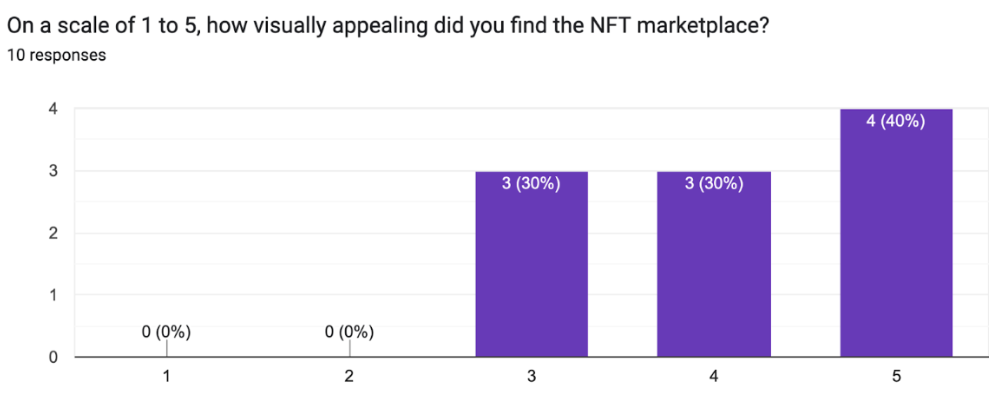}
    \caption{Usability study Q4 results}
    \label{fig: Q4}
\end{figure} 

 When it came to finding NFTs to purchase, 5 users found it very easy, 2 users found it somewhat easy, and 1 user found it difficult. The buying process was straightforward to use for 7 users, while 3 users rated it as very easy. In terms of the overall experience, 6 users rated it as good and 3 users rated it as very good. One user had a suggestion for improvement regarding the Metamask wallet not dynamically updating the address. 

\begin{figure}[!h]
    \centering
    \includegraphics[width=0.5\textwidth]{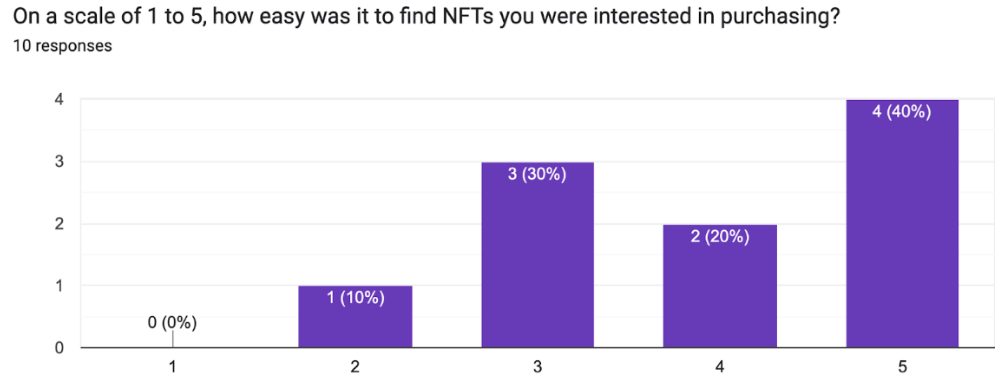}
    \caption{Usability study Q5 results}
    \label{fig: Q5}
\end{figure} 

\begin{figure}
    \centering
    \includegraphics[width=0.5\textwidth]{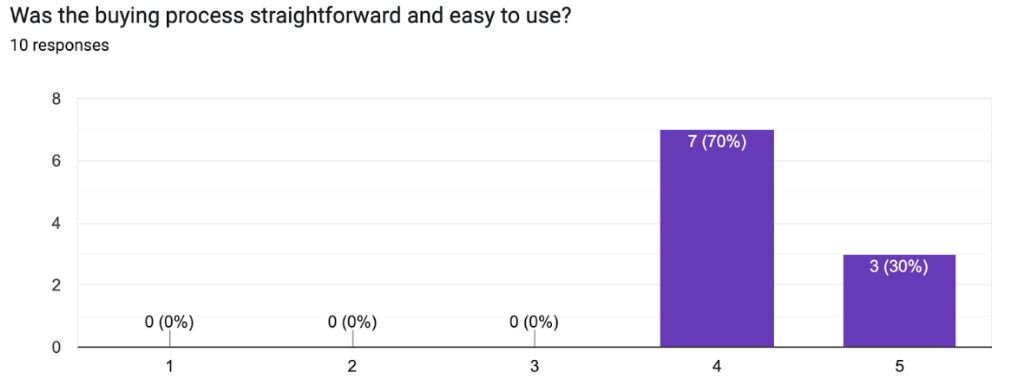}
    \caption{Usability study Q6 results}
    \label{fig: Q6}
\end{figure} 

\begin{figure}
    \centering
    \includegraphics[width=0.5\textwidth]{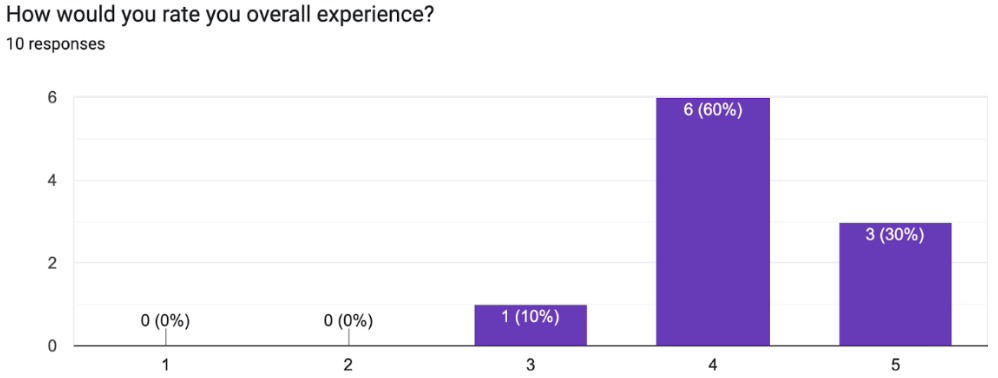}
    \caption{Usability study Q7 results}
    \label{fig: Q7}
\end{figure} 

\begin{figure}
    \centering
    \includegraphics[width=0.5\textwidth]{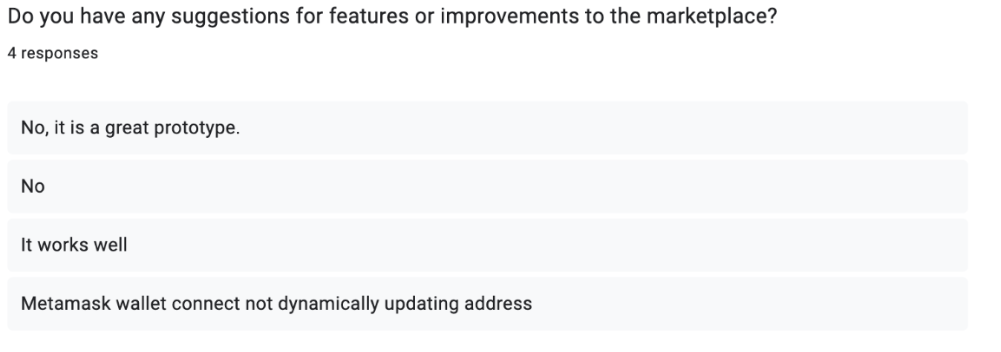}
    \caption{Usability study Q8 results}
    \label{fig: Q8}
\end{figure} 

\newpage

Based on our performance and usability evaluation, we have found our approach to be effective for generating and minting NFTs on the blockchain. Our dApp is user-friendly and accessible to users with varying levels of experience in blockchain-based applications. Nevertheless, it is important to note that more testing is necessary to assess the usability of our approach in different usage scenarios and with various user demographics. Additionally, further testing is required to evaluate the scalability and robustness of our approach under different network conditions and usage scenarios.

\section{Limitations and Future Work}
While the NFT Marketplace dApp offers a novel approach to secure digital asset management using blockchain technology and deep learning models, several limitations can be addressed in future iterations of the application.

Firstly, the NFT generation process is currently limited to the OpenAI API, which has several constraints, such as the maximum image resolution and the number of API requests allowed monthly. Future work could involve the development of custom deep learning models that are better suited to the generation of NFTs and can handle a larger variety of input types.

Secondly, the application is currently implemented on the Goerli testnet, which is a testing environment for Ethereum-based networks. While this allows us to test the application in a safe and controlled environment, it also limits the scalability and real-world applicability of the dApp. Future work could involve deploying the dApp on a mainnet to enable the trading of actual NFTs.

Thirdly, while the application currently supports the minting and trading of NFTs, it does not include support for the resale of NFTs. This is an important feature for NFT marketplaces, and future work could involve the development of a mechanism for the secure and transparent resale of NFTs.

Finally, the user interface of the dApp could be further improved to enhance the user experience and increase the accessibility of the platform to a wider audience. This could involve the use of more advanced design and usability principles, as well as the incorporation of additional features, such as social sharing and community-building tools.


\section*{Conclusion}

In conclusion, the development of the NFT Marketplace dApp demonstrates the potential of blockchain technology and deep learning models in creating secure, transparent, and user-friendly platforms for managing and trading digital assets. Our dApp addresses several fundamental challenges in NFT management, including secure wallet connection, NFT generation using deep learning, NFT minting on the blockchain, and a marketplace for trading NFTs. The integration of the OpenAI API and the Ethereum network has enabled the generation of personalized and unique NFTs that cannot be easily replicated, creating valuable and distinctive digital assets. The usability study conducted demonstrates that our dApp is user-friendly and easy to navigate. Future directions for research and development in this area may include improving the scalability of NFT marketplaces, exploring new approaches for NFT generation, and addressing potential security issues that may arise in the management and trading of digital assets. Overall, our NFT Marketplace application represents a significant step forward in the development of decentralized applications for NFT management and trading.



%

\end{document}